\documentclass[11pt,a4paper,reqno]{amsart}

\usepackage[utf8]{inputenc}
\usepackage[english]{babel}
\usepackage{amsmath}

\usepackage{amssymb, mathabx}
\usepackage[numbers]{natbib}
\usepackage{graphicx}

\usepackage{braket}
\usepackage[colorlinks,hyperindex,bookmarksopen,linkcolor=red,citecolor=blue,urlcolor=blue]{hyperref}

\usepackage{mathrsfs}

\usepackage{braket}

\bibpunct{[}{]}{;}{n}{,}{,}
\usepackage[hmargin=2cm,vmargin={3cm,4cm}]{geometry}

\theoremstyle{definition}                                 

\theoremstyle{definition}                           

\theoremstyle{remark}                             

\usepackage{color}

\usepackage{mathtools,slashed}

\newcommand{\be}{\begin{eqnarray}}
\newcommand{\ee}{\end{eqnarray}}

\numberwithin{equation}{section}

\allowdisplaybreaks

\usepackage{etoolbox}
\newcommand*{\affaddr}[1]{#1} 
\newcommand*{\affmark}[1][*]{\textsuperscript{#1}}

\begin{document}
\title{A transient biological fouling model for constant flux microfiltration}

\author{V.~Luongo\affmark[1]}

\author{M.R.~Mattei\affmark[1]}

\author{L.~Frunzo\affmark[1]}

\author{B.~D'Acunto\affmark[1]}

\author{K.~Gupta\affmark[3]}

\author{S.~Chellam\affmark[3,4]}

\author{N.G.~Cogan\affmark[2]}

\maketitle

{\footnotesize
  \begin{center}
\affaddr{\affmark[1]University of Naples "Federico II", Department of Mathematics and
Applications, \\ 
via Cintia, Monte S. Angelo I-80126 Napoli, Italy}\\
\affaddr{\affmark[2]Florida State University, Department of Mathematics, 208 Love Building, Tallahassee, FL 32306-4510, USA}\\
\affaddr{\affmark[3]Texas A\&M University, Department of Civil \& Environmental Engineering, College Station, TX 77843, USA}\\
\affaddr{\affmark[4]Texas A\&M University, Department of Chemical Engineering, College Station, TX 77843, USA}\\
   \end{center}}
	
	Corresponding author: V.~Luongo, \texttt{vincenzo.luongo@unina.it}\\

%
%
%

\begin{abstract}

Microfiltration technology is a widely used engineering strategy for fresh water production and water treatment. The major concern in many applications is the formation of a biological fouling layer leading to increased hydraulic resistance and flux decline during membrane operations. The growth of bacteria constituting such a biological layer implicates the formation of a multispecies biofilm and the consequent increase of operational costs for reactor management and cleaning procedures. To predict the biofilm growth and evolution during the filtration process, a one-dimensional continuous model has been developed by considering a free boundary value problem describing biofilm dynamics and EPS production in different operational phases of microfiltration systems. The growth of microbial species and EPS is governed by a system of hyperbolic PDEs. Substrates dynamics are modeled thorough parabolic equations accounting for diffusive and advective fluxes generated during the filtration process. The free boundary evolution depends on both microbial growth and detachment processes. The proposed model has been solved numerically to simulate biofilm evolution during biologically relevant conditions, and to investigate the hydraulic behavior of the membrane. The model has been calibrated and validated using lab scale experimental data. In all cases, numerical results accurately predicted the membrane pressure drop occurring in the microfiltration system.
	
\end{abstract}

\maketitle

\section{Introduction} \label{n1}

Membrane technology has been largely used as one of the most promising engineering strategies for both wastewater treatment and fresh water production \cite{Valladares_15}. Due to the increasing water demands for human, industrial, and agricultural use, many applications to provide clean water have been developed, primarily differentiated by membrane porosity, i.e. ultrafiltration, microfiltration, nanofiltration and reverse osmosis. The major concern in all these applications is the formation of a colloidal inorganic (e.g. scale) and/or biological (e.g. bacteria, organic particle flocs, bio-polymers) fouling layer \cite{Martin_14}. The latter is strongly influenced by the specific context as it depends on membrane characteristics and water quality. The accumulation of particles and the formation of a cake layer on membrane surfaces lead to increasing hydraulic resistance and productivity decline \cite{venezuela2009hybrid}. This represents a significant operational cost in terms of energy and chemicals for cleaning procedures \cite{Review_Anis19}.

Membrane performance is highly dependent on the characteristics of the treated water, e.g. macro and micro nutrient content, total and suspended solid content, and on the specific environment where the filtration process occurs. In the case of membrane bio-reactors (MBR), polymeric membranes for solid/liquid separation are used in bio-reactors where specific biological processes are catalyzed allowing for high quality effluents, low sludge production, and improved nutrient removal \cite{Ivanovic_2012}. For instance, organic carbon removal and nitrification process, operated by heterotrophic and autotrophic species, respectively, lead to the growth and accumulation of bacteria, which aggregate in active sludge flocs in aerated conditions and are able to ensure high quality of wastewater treatment plants effluents \cite{Mathematics_19, EJAM_18, trucchia}. Clearly, the direct contact of the membrane surface and bacterial flocs and microorganisms stimulates the formation of a mainly biological fouling layer constituting a multispecies biofilm.

Submerged membranes are bound to be colonized by bacteria \cite{Dreszer_2013}; they represent a perfect environment for biofilm growth and evolution. Initially, biofilms have a beneficial effect due to their ability to remove biodegradable pollutants \cite{Kang_2007}. Afterwords, they are responsible for an unacceptable decline of membrane performance, reduced water quality, and biodeterioration of membranes components \cite{Kerdi_2019,Dreszer_2013}. Over the last decades, mathematical modeling of membrane filtration systems have been largely studied using classical blocking laws \cite{chellamCompCake06,coganchell2009,brower95}. These are classified by the size of particulate foulants approaching the membrane and the size of pores constituting the filtration membrane. Two different approaches have been used for blocking law formulation: the constant flux $J$ approach, where increasing hydraulic resistance leads to increasing pressure drop $\Delta P(t)$; the constant pressure approach, where $J$ is a function of time and $\Delta P$ is a constant. These approaches reflect the conventional operation strategies usually adopted in membrane reactors. Despite the usefulness and accuracy of such models in diagnosing the hydraulic behavior both in forward operation and after backwashing, these models lack biological and kinetic description of fouling formation and development, as well as predictive ability.

Due to recent advances in mathematical modeling of multispecies biofilms and recent improvements in microscopy and imaging techniques \cite{Mattei_rev,Laspidou_14,tenore_21}, many researchers have begun to focus on the structural organization and rheological response of biofilms growing on membrane systems \cite{Jafari19,Dreszer_2013}. Vrouwenvelder \textit{et al.} \cite{Vrouw_16} assessed that the extracellular polymeric substances (EPS) constituting a biofilm exclusively determine its hydraulic resistance. In addition, the authors highlighted the negligible effect of bacterial cells embedded in the biofilm matrix. They proposed a 3-D mathematical formulation to predict the increase of hydraulic resistance during the permeation of water in a rectangular domain. Tierra et al. \cite{tierra15} introduced a phase-field multicomponent model to investigate the effect of EPS viscosity and elasticity on biofilm deformation. The authors were able to numerically reproduce the effect of the shear flow on the detachment rate by using flow cell experiments to determine the mechanical characteristics of the investigated biofilm. Recently, Li et al. \cite{li2020} proposed a phase-field continuum model coupled with the Oldroyd-B constitutive equation to simulate biofilm deformation under stress conditions. The model predicted with high accuracy the viscoelastic deformation of different mature biofilms constituting a useful tool for engineering biofilm systems control. However, all these models completely neglect the biological dynamics, the contribution of biofilm growth on fouling, and the related pressure drop occurring in membrane systems.

Other authors explored the mathematical modeling of membrane reactors exclusively focusing on the quality of effluents, or on the hydraulic response of biofilms under different stress conditions \cite{Tenore_18,Rahimi_09,Zare_13}. More sophisticated multidimensional models have been also proposed to describe the non-uniform development of particulate fouling layers and their heterogeneous morphology in high pressure cross-flow filtration membranes \cite{Picio_09,Shin_13}. Several of the authors here have studied the fouling and regeneration process, focusing on developing models that are amendable to optimal control analysis and extensions to more realistic models \cite{chellam2011colloidal,cogan2016optimal}.

To the best of our knowledge, mathematical models accounting for both detailed biofilm growth dynamics and EPS production during the operational phases of membrane reactors have not yet been developed. Therefore, this work connects insights related to membrane filtration, and growth and development of multispecies biofilms occurring in submerged filters. It aims at the development of a mathematical tool able to describe the filtration and backwashing effect on biofilm systems and the biofouling kinetics effect on the hydraulic behavior in microfiltration membrane systems. It represents a first step in the development of more complex mathematical models able to assist membrane facilities in designing and operational procedures.

The work includes observations of membrane/biofilm interaction, such as monodimensional spatial distribution of biofilm components, and substrate dynamics, and physical effects of backwashing on biofilm growth. The model was calibrated and validated by using experimental data obtained with a lab scale membrane system under different operating conditions (filtration time of $20$ and $40 \ min$). Numerical simulations remarked on the consistency of the model and showed the effect of substrate diffusion/convection and biofilm detachment during forward and backwashing operations.


 \section{Filtration principles and biofilm growth} \label{n2}

The microfiltration mathematical problem is here presented as a multi-scale model, where the pressure drop during water filtration (macroscale) is directly influenced by biofilm growth and development on the dead-end membrane surface (microscale). On the other hand, the effect of backwashing on biofilm dynamics, which affects the detachment rate and the free boundary problem describing biofilm thickness evolution, has been included in the present formulation.

\begin{figure}[h]
	\centering 
		\includegraphics[width=0.65\textwidth]{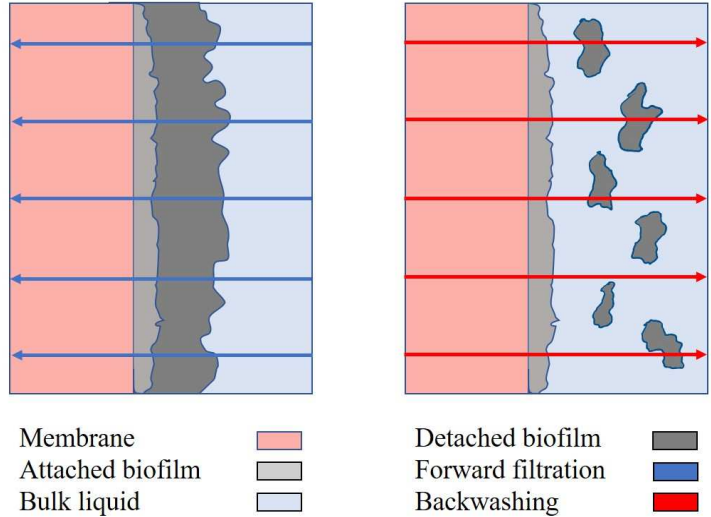}
	\caption{Schematic representation of MBR operational conditions: forward filtration (left) and backwashing operations (right).}
	\label{scheme}
\end{figure}

Figure \ref{scheme} resumes the operational conditions of the membrane system. During forward filtration, the wastewater is forced to pass through the membrane with consequent particles accumulation and biofilm growth stimulation. In the backwashing regimen, clean water is added to the system from the membrane surface to the bulk liquid as this practice partially removes previously formed cake layer including the biofilm.

Assuming constant flux $J$ operation during water filtration, the pressure drop $\Delta P(t)$ is defined by a Darcy's law based formulation as:

\begin{equation}						\label{d.1}
\Delta P(t) = J \mu \left(R_m + R_B(t)\right),
\end{equation}

\noindent
where $\mu$ is the absolute viscosity of water, $R_m$ is the hydraulic resistance of the clean membrane, and $R_B$ is the hydraulic resistance due to the presence of the biofilm layer. The constant flux assumption is motivated by engineering applications of the system. In Equation (\ref{d.1}), the hydraulic resistance of the biofilm layer $R_B$ is a function of the mass of foulants, i.e. biofilm components, accumulating on the membrane surface and constituting the biofouling cake.

The multispecies biofilm dynamics have been modeled as an essentially hyperbolic free boundary problem, and the biofilm thickness $L$ evolution, which represents the free boundary \cite{norwa}, is governed by the growth of biofilm components $X_i(z,t), \ i=1,...,n$, and the availability of substrates $S_j(z,t), \ j=1,...,m$ within the biofilm. Note that $X_i$ and $S_j$ are both concentrations $[mg L^{-1}]$, and each of the $i^{th}$ biofilm component can be expressed in terms of biofilm volume fraction $f_i$ as $X_i=\rho_i f_i, \ i=1,...,n$, where $\rho_i$ is the density in $[mg L^{-1}]$ of the $i^{th}$ biofilm component \cite{wanner86,invasion15}. These components are generally named biomasses and can be categorized as microbial species, which evolve over time due to their metabolic activities, and other biologically produced components, such as inert materials or extracellular polymeric substances (EPS), which are generated by living cells and accumulate in the biofilm matrix. For instance, the EPS fraction confers to the biofilm structure specific functions including the enhancement of the mechanical resistance \cite{laspidou2002non}.

The dynamics of biofilm components are derived from local mass balance considerations, and the hyperbolic system of Equations is described as:

\begin{equation}                                        \label{d.2}
			\frac{\partial X_i}{\partial t} +\frac{\partial}{\partial z}(u X_i)
			 =\rho_i r_{M,i}^{}(z,t,{\bf X},{\bf S}),\ \ i = 1,...,n,\ \ 0\leq z\leq L(t),\ t>0,
\end{equation}	

\noindent
where $u(z,t)$ is the velocity of microbial mass displacement with respect to the membrane surface, $r_{M,i}^{}(z,t,{\bf X},{\bf S})$ is the biomass growth rate, ${\bf X}=(X_i,...,X_n)$, and ${\bf S}=(S_j,...,S_m)$. The metabolic reactions operated by bacterial species are able to generate an advective flux described as a mass displacement, and dominated by microbial growth, decay reactions, and EPS production. The velocity of microbial mass displacement $u(z,t)$ is obtained by summing Equation (\ref{d.2}) over $i$,

\begin{equation}                                        \label{d.3}			
	     \frac{\partial u}{\partial z} = \sum_{i=1}^{n}r_{M,i}(z,t,\textbf{X},\textbf{S}), \ \ 0< z\leq L(t),\ t\geq0,
\end{equation}
 
\noindent
and considering the volume fraction constrain $\sum_{i=1}^n f_i=1$.
The multispecies biofilm growth model is completed by using an ordinary differential equation describing the evolution of the biofilm thickness $L(t)$ as a free boundary layer. The variation of biofilm thickness over time is described as:

\begin{equation}                                        \label{d.4}
	    \dot L(t)=u(L(t),t)-\sigma_d(L(t)), \ \ t>0,
\end{equation}

\noindent
which is a function of both the microbial mass displacement velocity and the detachment flux $\sigma_d(L(t))$. The latter is directly connected to membrane operation, as it accounts for the sloughing phenomenon occurring during backwashing operations. According to previous studies \cite{morghe99s, wanner86}, the proposed sloughing rate assumes that the loss of biofilm mass is proportional to the biofilm thickness. In addition, it can take into account that the EPS volume fraction is more resistant than all the other biofilm component to the hydraulic stress \cite{Vrouw_16}, and it negatively influences the efficiency of the cleaning procedure applied with the backwashing phase. The formulation of the detachment rate proposed in the present study is described as: 

\begin{equation}                                        \label{d.5}
			\sigma_d(L(t))= \lambda L^{2}+ K\left(\frac{\left|J\right|+J}{2}\right) \left(1-\hat{f}_{EPS}\right) (L-L_{lim}),
\end{equation}

\noindent
where, $J$ is the constant flux applied for the membrane filtration and the backwashing procedure, $\hat{f}_{EPS}$ is the average EPS volume fraction along the biofilm thickness, $K$ is the sloughing constant [$m^{-1}$], and $L_{lim}$ is an irreversible fouling layer that is not removed by backwashing procedures. Note that the second term on the right hand side of Equation (\ref{d.5}) is equal to zero when the filtration flux $J$ is non-positive, and this condition occurs when the water is filtered throughout the membrane surface or it is not filtered for MBR maintenance procedures. Thus the second term links the membrane operation and the foulant layer structure to the physical cleaning provided by backwashing.

The bacterial growth is catalyzed by the presence of $m$ different substrates in the treated wastewater, which are able to influence the metabolic activity of the considered microbial species constituting the biofilm. The evolution of substrates over time and space has been modeled as a system of nonlinear convection$-$diffusion$-$reaction equations:

\begin{equation}                                        \label{d.6}
	    \frac{\partial S_j}{\partial t}-\frac{\partial}{\partial z}\left(
   D_{S,j}\frac{\partial S_j}{\partial z}\right)+v\frac{\partial S_j}{\partial z}=
   r_{S,j}(z,t,{\bf X},{\bf S}), \ \ j=1,...,m, \ \ 0 < z < L(t), \ t>0,
\end{equation}

\noindent
where $D_{S,j}$ is the diffusivity coefficient, $v$ is the filtration velocity, $S_j(z,t)$ represents the concentration of the substrate $S_j, \ j=1,...,m$, and $r_{S,j}(z,t,{\bf X},{\bf S})$ is the conversion rate of each substrate $j$. Note that, in the present work, the filtration surface has been set to $1$, so the filtration velocity $v$ and the filtration flux $J$ assume the same values as they should.

Despite many simplistic membrane models describing the filtration mechanism as a process merely affected by particle sizes and membrane characteristics (e.g. blocking laws), the definition of $R_B$ in Equation (\ref{d.1}) allows to elucidate the relation between the biofouling kinetic evolution and the hydraulic (Darcy-based) resistance in in-series filtration processes. This aspect is crucial as in both real scale and lab scale experiments the hydraulic resistance of membrane reactors is strongly influenced by biological dynamics and by bacteria attached on the membrane surface. According to experimental evidences \cite{Vrouw_16}, various components constituting a biofouling layer can differently affect the pressure drop during microfiltration. Moreover, a specific hydraulic resistance $\alpha_i, \ i=1,...,n$ has been assumed for each biofilm component, so that the pressure drop during constant flow microfiltration can be assumed as a proportional function of the mass of each specific biofilm component constituting the biofouling layer. Equation (\ref{d.1}) can be rewritten in the form:

\begin{equation}						\label{d.16}
\Delta P(t) = J \mu \left(R_m + \sum^{n}_{i=1}{\alpha_{0,i} \int^{L(t)}_0 \frac{f_i \rho_i}{L_0} z dz}\right),
\end{equation}

\noindent
where $\alpha_{0,i}$, \ $f_i$, and $\rho_i$ are the specific resistance, the volume fraction, and the density of each $i^{th}$ biofilm component, respectively.
Note that, assuming a mono-species biofilm ($f_1=1$), Equation (\ref{d.16}) leads to

\begin{equation}						\label{d.17}
\Delta P(t) =\mu \left(R_m + \frac{\alpha_{1} \rho_1 L^{2}(t)}{2L_0}\right) J,
\end{equation}

\noindent
where the relation between the pressure drop $\Delta P(t)$ and the biofilm thickness $L(t)$ is highlighted. Note that this is a novel connection between the detailed resistance, calculated based on the dynamics of the fouling layer and the macroscopic fouling laws.


 \section{Initial-boundary conditions} \label{n2.2}

The forward and backwashing problem is treated by setting different initial-boundary conditions related to the systems of nonlinear partial differential Equations (\ref{d.2})-(\ref{d.4}) and (\ref{d.6}). For Equation (\ref{d.2}), the initial conditions

\begin{equation}                                        \label{d.2.2.1}
       X_i(z,0)= X_{i0}(z), \ \ i=1,...,n, \ \ 0\leq z\leq L_0,\\
\end{equation}

\noindent
have been prescribed, where $X_{i0}(z)$ are general positive functions associated to the different components constituting the initial biofilm structure. A no flux condition has been set on the substratum in Equation (\ref{d.3}) and the initial value for $L(t)$ has been defined in Equation (\ref{d.4}) as:

\begin{equation}                                        \label{d.2.2.2}			
			u(0,t)=0, \ \  t\geq0, \ \ L(0)= L_0,
\end{equation}

\noindent
where $L_0$ is a positive constant. The initial substrate concentration profiles $S_j(z,0)$ are directly affected from the substrate concentrations in the secondary treated wastewater, and are defined as

\begin{equation}                                        \label{d.2.2.3}
  		S_j(z,0)=S_{j0}(z), \ \ j=1,...,m, \ \ 0\leq z\leq L_0,
\end{equation}

\noindent
where $S_{j0}(z)$ are assigned positive functions. The forward filtration problem is solved by adopting the boundary conditions for $J<0$

\begin{equation}                                        \label{d.7}
  		\frac{\partial S_j}{\partial z}(0,t)=0, \ \ \frac{\partial S_j}{\partial z}(L,t)=\frac{v}{D_j} \left(S_{jL}(t)-S_{j}(L,t)\right), \ \	j=1,...,m, \ \ t>0,
\end{equation}

\noindent
where the functions $S_{jL}(t)$ are related to the substrate concentration in the wastewater. The diffusive flux at $z=0$ has been set to zero as the concentration value of substrates in the first section of the biofilm layer is equivalent to the concentration in the permeate flux (treated water). Moreover, the variation of substrates at $z=L(t)$ is described as a filtration flux, which is directly influenced by the concentration of substrates in the wastewater $S_{jL}(t)$, and by the filtration flux $J$. The forward filtration phase is followed by a zero flux phase $J=0$ required for membrane maintenance. In this phase, the boundary conditions for Equations (\ref{d.6}) assume the following form 

\begin{equation}                                        \label{dd.7}
  		\frac{\partial S_j}{\partial z}(0,t)=0, \ \ \frac{\partial S_j}{\partial z}(L,t)=0, \ \	j=1,...,m, \ t>0,
\end{equation}

\noindent
as the advective flux zero and the water is not fed to the MBR system from both the membrane sides. After a zero flux phase, the backwashing procedure begins with reversing the water flow $J>0$, and clean water enters in the system from the initial biofilm layer ($z=0$) to the moving boundary ($z=L(t)$). Robin-Neumann boundary conditions have been prescribed to solve the system of Equations (\ref{d.6}) under this condition

\begin{equation}                                        \label{d.8}
  		\frac{\partial S_j}{\partial z}(0,t)=\frac{v}{D_j} S{j}(0,t), \ \ \frac{\partial S_j}{\partial z}(L,t)=0, \ \ j=1,...,m, \ t>0.
\end{equation}

\noindent
Equation (\ref{d.8}) describes the substrate fluxes applied on the biofouling layer during the cleaning operation procedures. It is a common practice to use just water during backwashing to clean the membrane and mitigate the biofouling formation. For this reason, just an advective flux was adopted for the substrate concentrations at $z=0$. On the other hand, at $z=L$ a Neumann condition was set as it was assumed that the effect of substrates concentration in the bulk liquid $S_{jL}(t)$ is negligible with respect to the filtration flux $J$.


 \section{Model application} \label{n2.3}

The mathematical model was specified to simulate the heterotrophic-autotrophic competition for oxygen usually occurring in wastewater treatment. In this context, it is possible to remove nutrients from wastewaters by catalyzing the microbial metabolism of these species, and providing oxygen to the biological units of the wastewater treatment plant. This strategy leads to the effective decrease of nutrient concentrations in the water and provides a clean effluent with enhanced water quality. For the specific application, a constant density $\rho=\rho_i, \ i=1,...,n$ has been assumed for all the biofilm components, and $4$ different components have been considered $n=4$. The modeled microbial species, such as autotrophic $X_1$ and heterotrophic bacteria $X_2$, are able to produce two more biofilm components, i.e. inert material $X_3$ and EPS $X_4$, due to their growth and evolution driven by ammonium nitrogen and organic carbon uptake, respectively \cite{Mathematics_19,laspidou2002non}. Indeed, the bacterial growth is catalyzed by the uptake of $3$ different substrates, $m=3$, such as ammonium nitrogen $S_1$, organic carbon $S_2$, and dissolved oxygen $S_3$. The latter is required for both organic carbon removal and nitrification process, operated by the heterotrophic and autotrophic species, respectively.

The kinetic growth rate $r_{Mi}(z,t,\textbf{X},\textbf{S})$ for the biofilm components $X_1$, $X_2$, $X_3$, and $X_4$ are expressed as Monod-like kinetics and the growth and evolution of each component is dominated by the presence and availability of substrates in time and space. Moreover, they account for different biological mechanisms, such as endogenous respiration, EPS production, decay-inactivation, and biodegradability of microbial components, which usually are included in multispecies biofilm modeling. These lead to non-linear multiparameter expressions describing the dynamics of each biofilm component due to substrates utilization and metabolites production during biofilm evolution. For the bacterial components $X_1$ and $X_2$,

 \begin{equation}                                        \label{d.9}
							r_{M,1} = \left((1-k_1)K_{\max,1}\frac{S_2}{K_{1,2}+S_2}\frac{S_3}{K_{1,3}+S_3} 
							          - b_{1}F_1\frac{S_3}{K_{1,3}+S_3} 
												- (1-F_1)c_{1}\right)X_1, \\
 \end{equation}		
 
 \begin{equation}                                        \label{d.10}
							r_{M,2} = \left((1-k_2)K_{\max,2}\frac{S_1}{K_{2,1}+S_1}\frac{S_3}{K_{2,3}+S_3}
							          - b_{2}F_2\frac{S_3}{K_{2,3}+S_3}
												- (1-F_2)c_{2}\right)X_2, \\
 \end{equation}

\noindent
where $K_{\max,i}$ denotes the maximum net growth rate for biomass $i$, $k_i^{}$ is the coefficient associated to EPS formation, $K_{i,j}$ represents the affinity constant of substrate $j$ for biomass $i$, $b_{i}$ denotes the endogenous rate for biomass $i$, $c_{i}$ is the decay$-$inactivation rate for biomass $i$, and $F_i^{}$ represents the biodegradable fraction of biomass $i$. The latter represents a fraction of the microbial component that is converted in composite particulate material due to metabolic reactions. It allows the production in the biofilm matrix of an inert component $X_3$ whose growth rate is expressed as

\begin{equation}                                        \label{d.11}
							r_{M,3} = (1-F_1)c_{1}X_1+(1-F_2)c_{2}X_2. \\
 \end{equation}

\noindent
Similarly, the accumulation of the EPS component $X_4$ within the biofilm matrix is due to bacteria activities during their evolution. The terms $(1-k_i^{})$ in Equations (\ref{d.9}) and (\ref{d.10}) indicate that a fraction of available substrates are used by $X_1$ and $X_2$ microbial species for EPS formation during the metabolic reactions for biomass production. Moreover, the growth rate for the EPS component $X_4$ is defined as

\begin{equation}                                        \label{d.12}
							r_{M,4} = k_1K_{\max,1}\frac{S_2}{K_{1,2}+S_2}\frac{S_3}{K_{1,3}+S_3}X_1+k_2K_{\max,2}\frac{S_1}{K_{2,1}+S_1}\frac{S_3}{K_{2,3}+S_3}X_2. \\
 \end{equation}

The conversion rates $r_{S,j}(z,t,\textbf{X},\textbf{S}), \ j=1,2,3$, related to substrates utilization for metabolic activities are described as Monod-like kinetics, which define the consumption of ammonia $S_1$, organic carbon $S_2$ and oxygen $S_3$ by biofilm components. These are described as follows:

\begin{equation}                                        \label{d.13}
r_{{S,1}^{}} = -\frac{1}{Y_2}\left((1-k_2)K_{\max,2}\frac{S_1}{K_{2,1}+S_1}\frac{S_3}{K_{2,3}+S_3}\right)X_2,
\end{equation}
\begin{equation}                                        \label{d.14}
r_{{S,2}^{}} =  -\frac{1}{Y_1}\left((1-k_1)K_{\max,1}\frac{S_2}{K_{1,2}+S_2}\frac{S_3}{K_{1,3}+S_3}\right)X_1,
\end{equation}
\begin{equation}                                        \label{d.15}
\begin{gathered}
r_{{S,3}^{}}= -\frac{(1-Y_1)}{Y_1}\left((1-k_1)K_{\max,1}\frac{S_2}{K_{1,2}+S_2}\frac{S_3}{K_{1,3}+S_3}\right)X_1 \\
-\frac{(1-Y_2)}{Y_2}\left((1-k_2)K_{\max,2}\frac{S_1}{K_{2,1}+S_1}\frac{S_3}{K_{2,3}+S_3}\right)X_2 \\
- b_{m,1}F_1\frac{S_3}{K_{1,3}+S_3}X_1- b_{m,2}F_2\frac{S_3}{K_{2,3}+S_3}X_2,
\end{gathered}
\end{equation}
\noindent
where $Y_i$ denotes the yield for each biomass $i$.


 \section{Experimental set up and and numerical simulations} \label{n3}

The proposed mathematical model was used to reproduce lab scale experiments performed with $4$ polyvinylidene fluoride (PVDF) hollow fiber membranes operated at a constant flux of $60 \ L m^{-2} h^{-1}$ during forward microfiltration. The total surface of the membrane modules was $30 \ cm^2$ and a secondary treated wastewater was fed to the system with a dissolved organic carbon (DOC) content around $6.2\pm0.12 \ mg L^{-1}$, corresponding to a chemical oxygen demand (COD) content of about $25 \ mg L^{-1}$ \cite{TOC_COD}. Different timing for forward filtration were used in two experimental sets, where the membranes were backwashed at $20$ or $40$ minutes intervals. Moreover, the same backwashing procedure was adopted in both the experimental sets. It consists in a $90 \ s$ flushing phase to remove the permeate and fill the tubes with ultra-pure water, followed by a $60 \ s$ backwashing step with ultra-pure water at the same working flux $J$. In addition, a $30 \ s$ gravity drain was applied, in order to completely empty the membrane modules, and a $100 \ s$ forward flushing phase was applied to fill the membrane modules with the feed wastewater, prior to start with the further filtration cycle. In all experimental cases, the total specific filtered volume was $720 \ L m^{-2}$. 

Virgin and fouled membranes were stained with two fluorescent dyes simultaneously: 4',6-diamidino-2-phenylindole dihydrochloride (DAPI), and 5-cyano-2,3-ditolyl tetrazolium chloride (CTC) (Biotium, CA) to observe total and live bacterial cells under the microscope, respectively \cite{number1}. DAPI fluoresces upon binding to DNA \cite{number1}, whereas CTC dye is reduced and fluoresces when there is an electron transport, implying actively respiring bacteria \cite{number2}. Stained samples were imaged using an Olympus BX53 microscope. The backwashing procedure was modeled with a zero flux phase of $90 \ s$ followed by a reversed flux phase of $60 \ s$. 
The biofilm-membrane model was able to reproduce the cleaning procedure with a zero flux $J=0$ and a reversed flux $J=60 \ L m^{-2} h^{-1}$ phase prior to start a new forward filtration $J=-60 \ L m^{-2} h^{-1}$ cycle.

\begin{figure}[h]
	\centering 
		\includegraphics[width=0.85\textwidth]{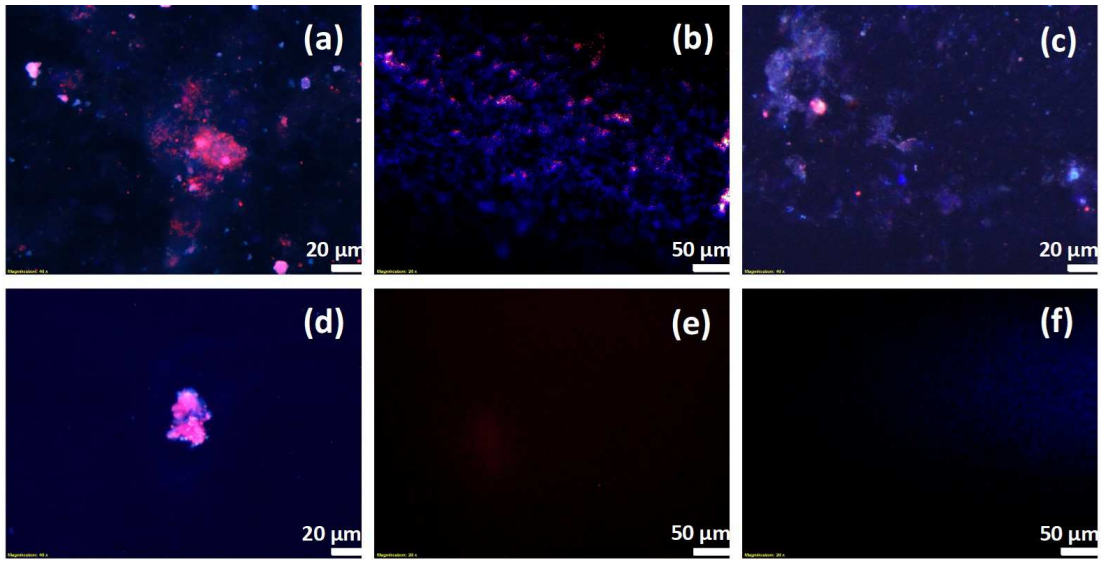}
	\caption{Optical microscope images of (a, b, c, and d) fouled membranes showing both blue and red fluorescence, and (e and f) virgin membrane showing neither red (CTC) or blue (DAPI) fluorescence.}
	\label{microscope}
\end{figure}

\noindent
As highlighted in Figure \ref{microscope}, fouled membranes showed the presence of both DNA (blue stain from DAPI) and as well as actively respiring bacteria (red stain from CTC). Higher intensity of red fluorescence obfuscated the blue fluorescence in some regions of the membrane, for example, Figure \ref{microscope}a and d, shows the typical clustering of live bacteria. In contrast, blue fluorescence was more spread out indicating the existence of biofilms covering the membrane surface. Besides, the virgin membrane showed no fluorescence, Figure \ref{microscope}e and f, serving as a control and confirming the buildup of bacteria and biofilm in the form of fouling during filtration.
To completely reproduce the experimental set-up, the numerical simulations were carried out without ammonium nitrogen, $S_1=0$ in the general mathematical model formulation, and the model was run by fixing a negligible initial autotrophic biofilm fraction $X_1=0.001$. In such specific situation, only heterotrophic bacteria can constitute the biofouling layer as their growth and evolution only require organic carbon $S_2$ and dissolved oxygen $S_3$.

The first data set, i.e. forward filtration for $20 \ min$, was used to calibrate the model. Based on the characteristics of the fed wastewater, the concentration of dissolved compounds in the bulk liquid, such as ammonia $S_1$, organic carbon $S_2$, and oxygen $S_3$, was fixed at $0$, $25$, and $8 \ mg L^{-1}$, respectively.
The initial biofilm thickness was set to $50 \ \mu m$, as it was assumed that a thin cake layer immediately appears on the membrane surface due to the presence of suspended bacteria and particles in the wastewater. The initial biofilm composition was characterized by a predominant heterotrophic bacteria component, whose activity is stimulated by the presence of organic carbon $S_2$ and dissolved oxygen $S_3$ under non-limiting conditions. The adopted initial autotrophic bacteria $f_2$ and inert materials $f_3$ volume fractions were $0.001$ and $0.002 \%$, respectively. The EPS volume fraction $f_4$, which usually depends on biofilm maturation and polymers production within the matrix during biofilm growth, was initially set to $0.05 \%$. The biofilm component growth rates and kinetic constants were derived from previous studies \cite{Mathematics_19,laspidou2002non,norwa}. Table \ref{t1} resumes all the kinetic constants and parameters adopted for numerical simulations. 

\begin{table}[!htb]
\begin{footnotesize}
\begin{center}
\caption{Kinetic parameters used for model simulations.} \label{t1}
\centering
\begin{tabular}{llcc}
\hline
{\textbf{Parameter}} & {\textbf{Definition}} & {\textbf{Unit}} &  \textbf{Value} \\
 \hline
 $K_{max,1}$   & Maximum growth rate for $X_1$                &  $d^{-1}$                              & $4.8$       \\
 $K_{max,2}$   & Maximum growth rate for $X_2$                &  $d^{-1}$                              & $0.95$      \\
 $k_1$         & EPS formation by $X_1$                       &  $mg COD/mg COD$                       & $0.45$      \\
 $k_2$         & EPS formation by $X_2$                       &  $mg COD/mg COD$                       & $0.34$     \\
 $K_{1,2}$     & Organics half saturation constant for $X_1$  &  $mg COD l^{-1}$                       & $5$         \\
 $K_{1,3}$     & Oxygen half saturation constant for $X_1$    &  $mg l^{-1}$                           & $0.1$       \\
 $K_{2,1}$     & Ammonium half saturation constant for $X_2$  &  $mg N l^{-1}$                         & $1$         \\
 $K_{2,3}$     & Oxygen half saturation constant for $X_2$    &  $mg l^{-1}$                           & $0.1$       \\
 $b_{1}$       & Endogenous rate for $X_1$                    &  $d^{-1}$                              & $0.025$     \\
 $b_{2}$       & Endogenous rate for $X_2$                    &  $d^{-1}$                              & $0.0625$    \\
 $F_{1}$       & Biodegradable fraction of $X_1$              &  $--$                                  & $0.8$       \\
 $F_{2}$       & Biodegradable fraction of $X_2$              &  $--$                                  & $0.8$       \\
 $c_{1}$       & Decay-inactivation rate for $X_1$            &  $d^{-1}$                              & $0.05$      \\
 $c_{2}$       & Decay-inactivation rate for $X_2$            &  $d^{-1}$                              & $0.05$      \\
 $Y_1$         & Yield of $X_1$                               &  ${g_{biomass}}/{g_{substrate}}$       & $0.4$       \\ 
 $Y_2$         & Yield of $X_2$                               &  ${g_{biomass}}/{g_{substrate}}$       & $0.22$      \\
 $\mu$         & Absolute viscosity of water                  &  $N s m^{-1}$                          & $10^{-3}$   \\
 $\rho$        & Biofilm components density                   &  $g m^{-3}$                            & $2500$      \\
 $\lambda$     & Biomass shear constant                       &  $m^{-1} d^{-1}$                       & $1250$      \\
 $K$           & Sloughing constant                           &  $m^{-1}$                              & $55.5$      \\
 \hline
 \end{tabular}
 \end{center}
 \end{footnotesize}
 \end{table}

The model was able to reproduce the biofouling formation over time during forward and backwashing operations of the system. The biofilm growth rate was higher during forward filtration as the availability of substrates for the microbial species was higher than in the backwashing step. Indeed, the convection and diffusion of substrates within the biofilm was strongly influenced by boudary conditions, as it is shown in Figure \ref{f1}:

\begin{figure}[h]
	\centering
		\includegraphics[width=0.75\textwidth]{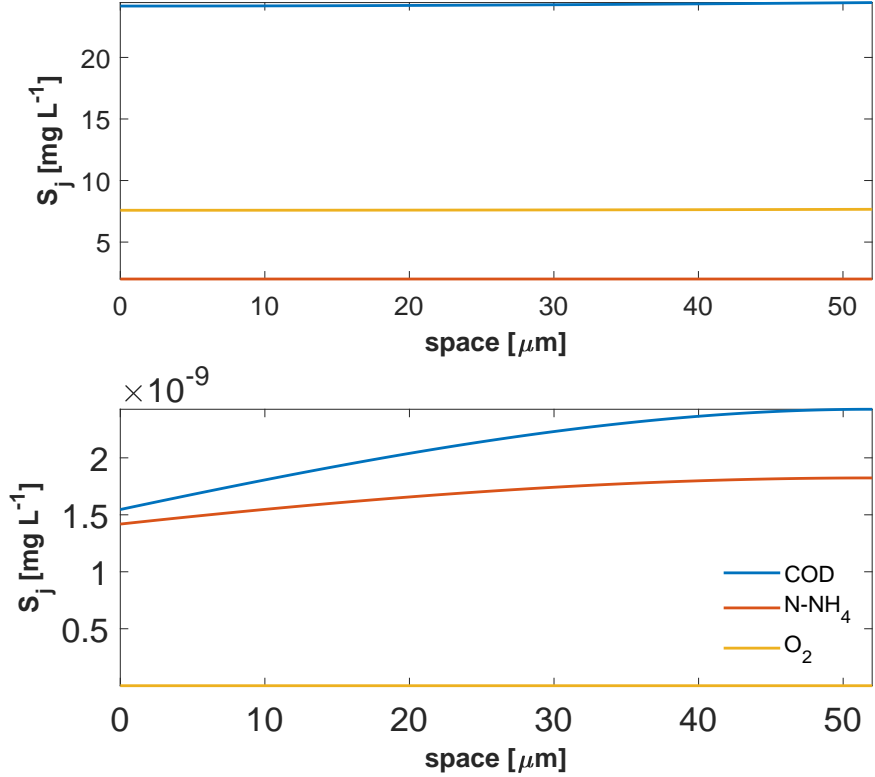}
	\caption{Substrate profiles during forward filtration (up) and backwashing operations (down). Note the y-axis scaling to low values.}
		\label{f1}
\end{figure}

\noindent
during forward filtration, the biofilm profile results fully penetrated by dissolved substrates, which lead to higher kinetic growth rates than during the other operating conditions with relatively high $COD$ concentration in the permeate water (Figure \ref{f1}a). Conversely, pure water crosses the biofilm from the membrane surface to the boundary layer during backwashing, and the dissolved substrates are completely washed out from the biofilm layer. Figure \ref{f1}b shows the extremely low substrate concentrations at the end of the backwashing phase.

For the newly introduced parameters, such as the hydraulic resistance of the membrane $R_M$, the specific resistance terms of each biofilm component $\alpha_i, i=1,...,n$, and the permanent biofouling layer $L_{lim}$, the values were obtained by comparing the model output and the experimental pressure drop data over time. The procedure will be explained in the next subsection. Table \ref{t2} resumes the adopted values for the initial condition of biofilm components and for substrate concentrations.

\begin{table}[!htb]
\begin{footnotesize}
\begin{center}
\caption{Initial conditions for biofilm growth.} \label{t2}
\centering
\begin{tabular}{lclcc}
\hline
{\textbf{Parameter}} & {\textbf{Symbol}} & {\textbf{Unit}} &  \textbf{Value} \\
 \hline
 
 Ammonia concentration    at $L=L(t)$                 &  $S_{1L}$       &  $mgl^{-1}$       & $0$     \\
 Organic Carbon concentration  at $L=L(t)$            &  $S_{2L}$       &  $mgl^{-1}$       & $25$    \\
 Dissolved Oxygen concentration  at $L=L(t)$          &  $S_{3L}$       &  $mgl^{-1}$       & $8$     \\
 Initial Biofilm thickness                            &  $L_0$          & $mm$              & $0.05$   \\
 Initial Volume Fraction of Heterotrophs $(X_1)$      &  $f_{1,0}(z)$   & --                & $0.947$  \\
 Initial Volume Fraction of Autotrophs $(X_2)$        &  $f_{2,0}(z)$   & --                & $0.001$  \\
 Initial Volume Fraction of Inert $(X_3)$             &  $f_{3,0}(z)$   & --                & $0.002$  \\
 Initial Volume Fraction of EPS $(X_4)$               &  $f_{4,0}(z)$   & --                & $0.05$   \\
 \hline
 \end{tabular}
 \end{center}
 \end{footnotesize}
 \end{table}

 \section{Calibration and validation} \label{n6}

The hydraulic resistance of the membrane filter $R_M$ and the specific hydraulic resistance values of each biofilm component $\alpha_i, i=1,...,n$ were obtained from the calibration procedure. The inclusion of $R_M$ in the calibration protocol was due to the highly different values reported in the literature for similar microfiltration systems. The biological components, such as autotrophic $X_1$ and heterotrophic $X_2$ bacteria, were supposed to react in the same way to the hydraulic filtration stress. Therefore, their specific hydraulic resistance $\alpha_1$ and $\alpha_2$ were constrained to have the same value, i.e. $\alpha_1=\alpha_2$. Similarly, the specific hydraulic resistance of the biologically produced components, such as inert material $X_3$ and EPS $X_4$, were constrained to have the same value, i.e. $\alpha_3=\alpha_4$. Moreover, different permanent fouling layers $L_{lim}$ were tested during the calibration. This parameter was added to the calibration procedure as Equation (\ref{d.5}) has been newly introduced in the present work to account for backwashing procedures occurring in wastewater systems. Indeed, the detachment rate $\sigma_{d}$ assumes an important rule in modeling of biofilm growth, as it usually dominates the dynamics of biofilm components for long-term behavior of the multispecies biological system \cite{Abbas_Eberl_2012}. In addition, the detachment occurring during backwashing procedures also affects the short-term filtration performance containing the pressure drop during forward filtration \cite{Guo_Ngo_2012}. Experimental evidences demonstrated the presence of a permanent fouling layer which cannot be removed by traditional backwashing procedures using clean water \cite{Mansouri_2010}. In same cases, the use of specific chemical cleaning protocols represent the only solution to restore the membrane, and completely eradicate the permanent biofouling layer. Moreover, aggressive cleaning procedures can seriously damage the filtration membrane, shortening its operating life \cite{Mansouri_2010}. In the present case, a simple water cleaning procedures was adopted and a permanent fouling layer $L_{lim}$ in the range of $0.01$ and $0.05 \ mm$ was tested with numerical experiments.

The Matlab tool \textsl{fmincons} was used to find the best set of parameters ($R_M, \alpha_i, i=1,...,4$) and minimize the Euclidean norm (EN) between model prediction and experimental data. Equation (\ref{n2n}) reports the discrete formulation of the minimized function used for the $\Delta P$-based calibration:

\begin{equation}						\label{n2n}
EN=\sum^{\bar{k}}_{k=1}{\sqrt{\left[J_k \mu \left(R_M + \rho\frac{L_k^2}{2 L_0} \sum^{4}_{i=1}{\alpha_{0,i} \ f_{i,k}} \right) \right]^2 - \bar{\Delta P_{k}}^2}}, \ \ k=1,...,\bar{k},
\end{equation}

\noindent
where the subscript $k$ represents a specific measurement/sampling time, $\bar{\Delta P_{k}}$ is the observed experimental value at the specific sampling time, and $\bar{k}$ is the total number of samples. The procedure was repeated for each permanent biofouling layer $L_{lim}$ value, i.e. $0.01$, $0.02$, $0.03$, $0.04$, and $0.05$, and the mean absolute relative error $\bar{\epsilon}$, Equation (\ref{e2e}), was used to compare model predictions with experimental data:

\begin{equation}						\label{e2e}
\bar{\epsilon}=\frac{1}{\bar{k}} \sum^{\bar{k}}_{k=1} \left|\frac{\Delta P_{m,k} - \bar{\Delta P_{k}}} {\bar{\Delta P_{k}}} \right| 100, \ \ k=1,...,\bar{k},
\end{equation}

\noindent
where $\Delta P_{m,k}$ represent the model prediction at a specific sampling time. Note that the term $\Delta P_{m,k}$ in Equation (\ref{e2e}) corresponds to $J_k \mu \left(R_M + \rho \frac{L_k^2}{2 L_0} \sum^{4}_{i=1}{\alpha_{0,i} \ f_{i,k}} \right)$ in Equation (\ref{n2n}).

\begin{figure}[!h]
	\centering
		\includegraphics[width=1\textwidth]{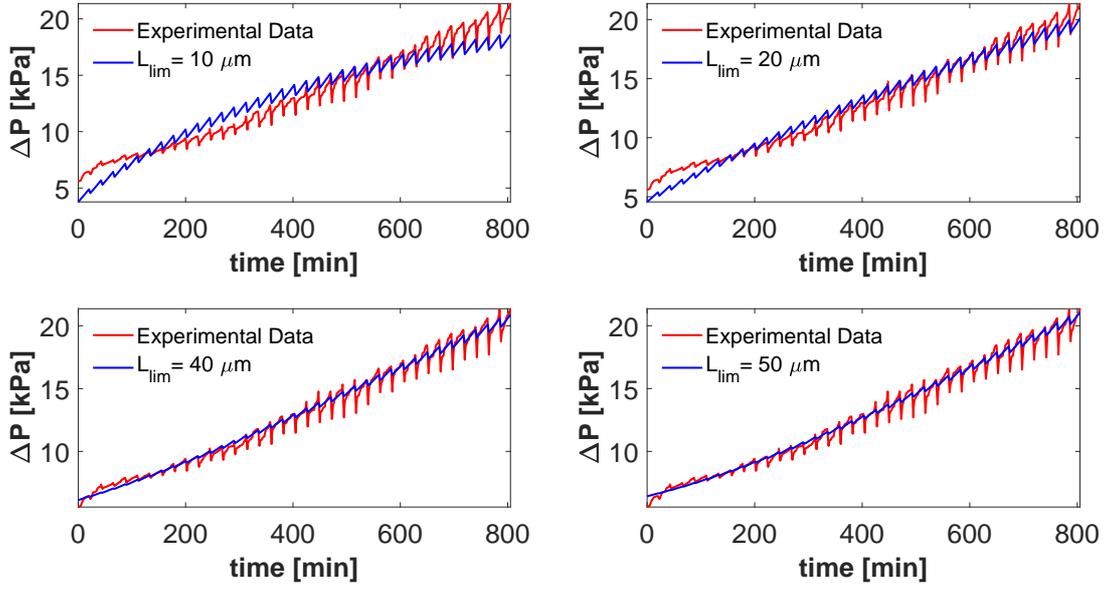}
	\caption{Numerical simulation with different permanent biofouling layer of $0.01 \ mm$, $0.02 \ mm$, $0.04 \ mm$, and $0.05 \ mm$. The observed values of the mean absolute relative errors are $8.56$, $4.98$, $2.15$, and $1.99 \%$, respectively.}
		\label{f5}
\end{figure}

The numerical analysis revealed a better fitting with experimental data when using higher values of $L_{lim}$. The higher was the permanent biofouling layer value, the lower mean absolute relative error $\bar{\epsilon}$ was observed, Figure \ref{f5}.

\begin{figure}[!h]
	\centering
		\includegraphics[width=0.9\textwidth]{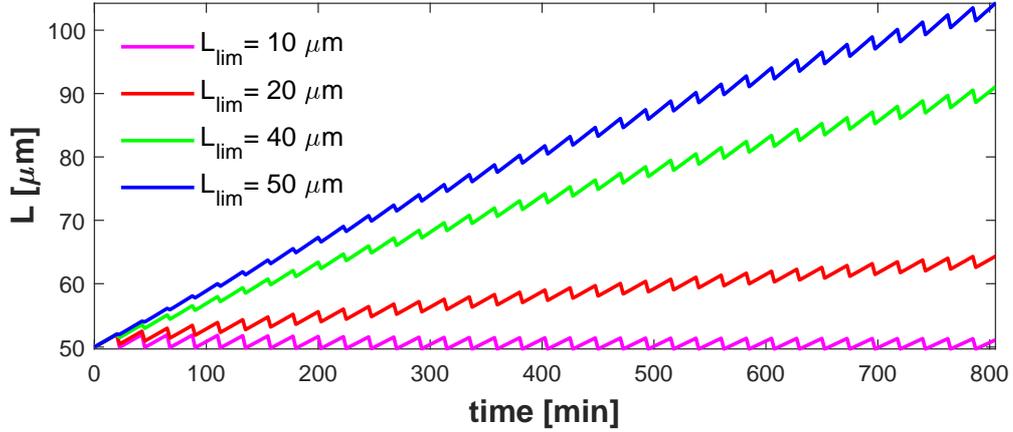}
	\caption{Biofilm thickness evolution using different $L_{lim}$ values of $0.01 \ mm$, $0.02 \ mm$, $0.04 \ mm$, and $0.05 \ mm$.}
		\label{f7}
\end{figure}

\noindent
This result was due to the strong influence of the permanent biofouling layer on the detachment rate $\sigma_d$, Equation (\ref{d.5}): when a low value of $L_{lim}$ is applied, numerical simulations showed a lower accuracy compared with higher values. In particular, the model was not able to fit the increasing pressure drop occurring during the first part of the data set. Different values of $L_{lim}$ led to different biofilm thickness profile over time. Figure \ref{f7} shows that the selection of the permanent fouling layer value is crucial to obtain a meaningful response from numerical simulations. Experimental evidences demonstrated a significant accumulation of biofilm on the top of the membrane, which cannot be reproduced using low biofouling layer values ($L_{lim}=10 \ \mu m$ and $L_{lim}=20 \ \mu m$). 

\begin{figure}[!h]
	\centering
		\includegraphics[width=0.75\textwidth]{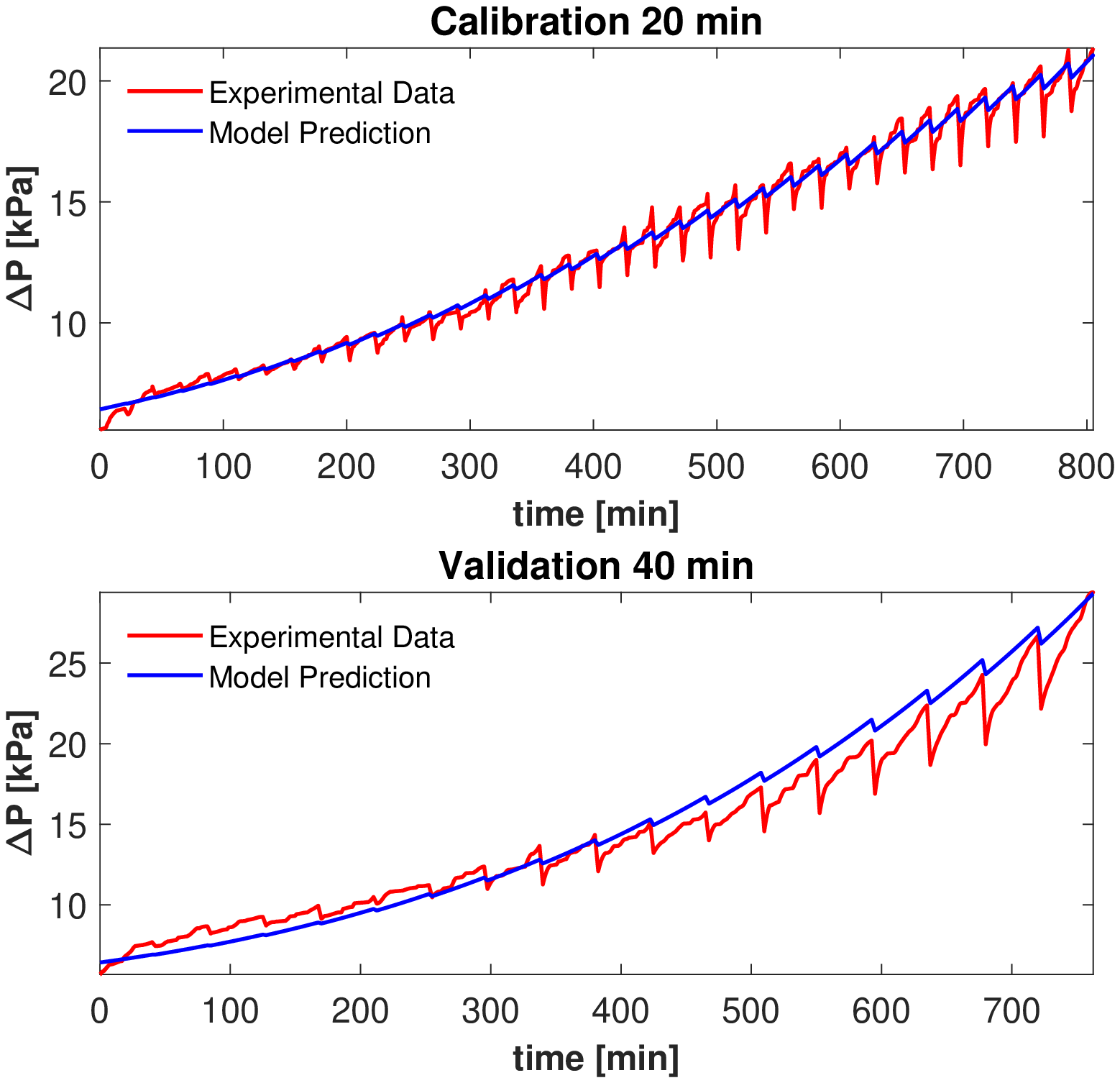}
	\caption{Model calibration with $20 \ min$ forward filtration (up, $\bar{\epsilon}=1.99 \%$) and model validation with $40 \ min$ forward filtration (down, $\bar{\epsilon}=7.07 \%$).}
		\label{f3}
\end{figure}

The best fit with experimental data $L_{lim} = 50 \ \mu m$ was used for model calibration using the $20 \ min$ filtration dataset. The mean absolute relative error $\bar{\epsilon}=1.99 \%$ was achieved and the calibrated values ($R_M, \alpha_i, i=1,...,4$, and $L_{lim}$) were used for model validation. Indeed, using the $40 \ min$ filtration dataset the model showed a mean absolute relative error $\bar{\epsilon}=7.07 \%$ as reported in Figure \ref{f3}. 

\noindent
The calibrated parameters are summarized in Table \ref{t3}.

\begin{table}[!htb]
\begin{footnotesize}
\begin{center}
\caption{Calibrated parameters.} \label{t3}
\centering
\begin{tabular}{lclcc}
\hline
{\textbf{Parameter}} & {\textbf{Symbol}} & {\textbf{Unit}} &  \textbf{Value} \\
 \hline
 
 Resistance of the clean membrane                       &  $R_M$          &  $mm^{-1}$          & $3.29\cdot10^{5}$  \\
 Specific resistance of heterotrophic bacteria $(X_1)$  &  $\alpha_1$     &  $mm Kg^{-1}$       & $2.20\cdot10^{5}$  \\
 Specific resistance of autotrophic bacteria $(X_2)$    &  $\alpha_2$     &  $mm Kg^{-1}$       & $2.20\cdot10^{5}$  \\
 Specific resistance of inert materials $(X_3)$         &  $\alpha_3$     &  $mm Kg^{-1}$       & $1.34\cdot10^{7}$  \\
 Specific resistance of EPS $(X_4)$                     &  $\alpha_4$     &  $mm Kg^{-1}$       & $1.34\cdot10^{7}$  \\
 Permanent biofouling layer                             &  $L_{lim}$      &  $mm$               & $0.05$             \\

 \hline
 \end{tabular}
 \end{center}
 \end{footnotesize}
 \end{table}
 
\noindent

Noteworthy, the specific hydraulic resistance related to the biologically produced components is one order magnitude higher than the bacterial species resistance. This is in accordance with the experimental evidence demonstrating that the hydraulic resistance of the whole biofouling layer can be attributed to the EPS formation during biofilm growth \cite{Cog_Chel_14,Vrouw_16}. 

The model was able to fit experimental data and predict with good accuracy the pressure drop occurring in the last part of the experiments. It can be noticed that the model underestimates the real pressure drop occurring from $t=0$ to $t=350 \ min$, while an opposite trend was observed during the last part of the experiment. This is a crucial aspect as the higher is the pressure levels needed in the membrane-biofilm system, the higher are operation costs occurring in real scale applications. This observation confers more relevance to higher pressure levels occurring during biofilm maturation, $t>350 \ min$, than to the lower pressure drop characterizing the beginning of the experiment.


 \section{Conclusions} \label{n9}

A 1-D mathematical model describing the formation and evolution of biofilm in a microfiltration membrane system has been presented, motivated by the importance and economic impact of biofouling in microfiltration. The biological process of EPS production and accumulation in the biofilm matrix, and its influence on the increasing hydraulic resistance during filtration has been addressed. The present work represents a direct connection between the conventional hydraulic modeling of membrane filtration and the mathematical modeling of multispecies biofilm growth and dynamics. A novel formulation of the detachment rate has been introduced. It is able to account for the effect of backwashing procedures on biofilm development. The interaction of heterotrophic and autotrophic species occurring in wastewater treatment units has been considered. To this aim, the free-boundary problem of biofilm growth on a filtration support has been numerically solved using the method of characteristics. The model accounts for substrate diffusion/convection and its effect on biofilm growth.

The model was calibrated and validated by using lab-scale experimental data of microfiltration. It was able to predict in a reasonable way the increase of pressure levels during constant flux microfiltration. Future studies are still required to address specific biofilm and membrane behaviors occurring in larger scale applications for wastewater treatment. Numerical simulations confirmed the crucial role of the biofilm EPS matrix on membrane pressure drop. A next step might be the application of the model to different hydraulic regimes and different biological cases occurring in engineering water filtration systems. 


 \section{Acknowledgements}\label{n5}

The authors also acknowledges the support from: CARIPLO Foundation (progetto VOLAC, Grant number: 2017-0977); Progetto Giovani G.N.F.M. 2019 "Modellazione ed analisi di sistemi microbici complessi: applicazione ai biofilm". S. Chellam gratefully acknowledges funding from the National Science Foundation (CBET 1636104).

This paper has been performed under the auspices of the G.N.F.M. of I.N.d.A.M. 

 \section{Appendix A - Other model modifications} \label{n8}

\begin{figure}[!h]
	\centering
		\includegraphics[width=1\textwidth]{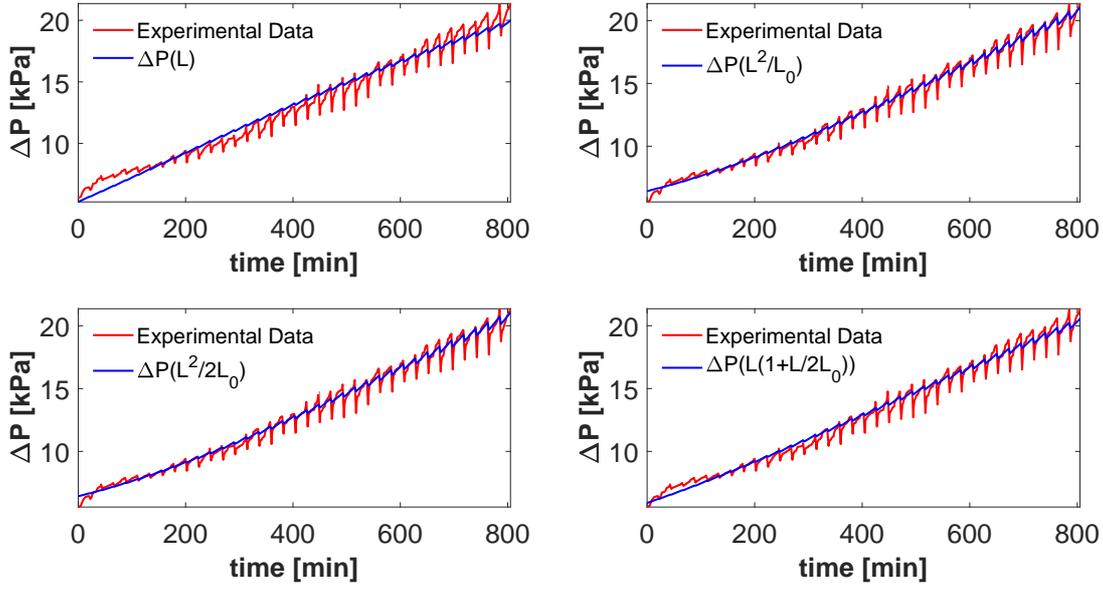}
	\caption{Numerical simulation with different $\Delta P(t)$ formulations. The relative errors obtained using Equations (\ref{d.19})-(\ref{d.22}), are $4.22$, $1.99$, $1.99$, and $2.67 \%$, respectively.}
		\label{f6}
\end{figure}

The newly introduced formulation for the biofilm hydraulic resistance, $R_B(t)$ in Equation (\ref{d.1}) and (\ref{d.16}), was obtained by comparing the real available experimental data with different theoretical formulations of the pressure drop occurring during MBR operations. Linear (Equation (\ref{d.19})) and quadratic (Equations (\ref{d.20}), (\ref{d.21}) and (\ref{d.22})) correlations between the pressure drop $\Delta P$ and the biofilm thickness were derived from mass balance principles, and the following formulations 

\begin{equation}						\label{d.19}
\Delta P(t) = J \mu \left(R_m + \sum^{n}_{i=1}{\alpha_{0,i} \int^{L(t)}_0 f_i \rho_i dz}\right),
\end{equation}

\begin{equation}						\label{d.20}
\Delta P(t) = J \mu \left(R_m + \sum^{n}_{i=1}{\alpha_{0,i} \int^{L(t)}_0 \frac{2 f_i \rho_i}{L_0} z dz}\right),
\end{equation}

\begin{equation}						\label{d.21}
\Delta P(t) = J \mu \left(R_m + \sum^{n}_{i=1}{\alpha_{0,i} \int^{L(t)}_0 \frac{f_i \rho_i}{L_0} z dz}\right),
\end{equation}

\begin{equation}						\label{d.22}
\Delta P(t) = J \mu \left(R_m + \sum^{n}_{i=1}{\alpha_{0,i} \int^{L(t)}_0 f_i \rho_i \left(1+\frac{z}{L_0}\right) dz}\right).
\end{equation}

\noindent
were tested using the same minimizing tool \textsl{fmincons} to compare model predictions and lab scale data in all the performed numerical experiments. This means that the different formulations presented in Equations \ref{d.19}-\ref{d.22} were used during the calibration and validation steps and when changing the model initial conditions. An typical result is shown in Figure \ref{f6}, where the minimum mean average errors, were obtained by using Equation (\ref{d.20}) and (\ref{d.21}). Indeed, these formulations are quite similar as they differently consider the effect of the hydraulic resistance of biological and non-biological components by scaling the values of the parameters $\alpha_{0,i}, i=1,...,4$ by a factor of $2$. The Equation (\ref{d.21}) was then used for the general model formulation, as in some cases the mean average relative error was lower than the observed value obtained by using Equation (\ref{d.20}).

\begin{figure}[!h]
	\centering
		\includegraphics[width=0.75\textwidth]{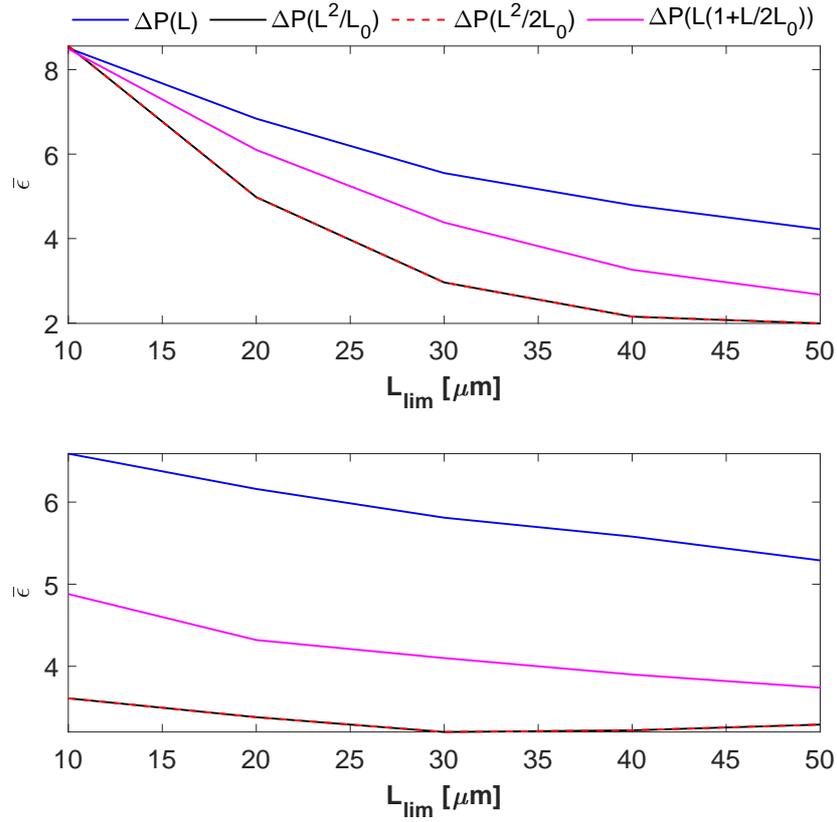}
	\caption{Mean average relative error trends with different $\Delta P(t)$ formulations obtained with $20 \ min$ (top), and $40 \ min$ (bottom) forward filtration.}
		\label{f8}
\end{figure}

A quadratic correlation of the pressure drop $\Delta P$ and the biofilm thickness allows for a more reasonable data fitting, and Equations \ref{d.20} and \ref{d.21} showed the best results. Indeed, Equations (\ref{d.19})-(\ref{d.22}) were individually used to determine the pressure drop in all the numerical experiments performed during the calibration and validation phases. For instance, the mean average relative error trend related to each simulation set with increasing $L_{lim}$ was also analyzed to test the accuracy of the matematical model in fitting the pressure drop experimental data. Figure \ref{f8} shows the trends of the error occurring when increasing the permanent biofouling layer from $0.01$ to $0.05 \ mm$. The $20 \ min$ and $40 \ min$ data-sets were used as example.

\noindent
It can be notice that with $20 \ min$ forward filtration the error shows a fast decreasing trend when increasing the permanent biofouling layer, exactly as the second order norm calculated for the optimization. When $40 \ min$ forward filtration data are used, the minimum mean average error is higher then the one obtained in the first case, and it shows a slight increase when increasing the $L_{lim}$ value, Figure \ref{f8}. This behavior can be attributed to the more difficult prediction of the effective pressure drop and biofilm thickness when increasing the filtration time. In this case, the effect of the sloughing is more impulsive and the model results are less accurate than in the case of limited forward filtration time. Further experiment are still required to describe the error trends when higher values of $L_{lim}$ are used for the detachment rate.

The effect of the erosion term $\lambda L^{2}$ in the newly introduced detachment Equation (\ref{d.5}) was also investigated. Numerical simulations were run with fixing $\lambda=0$ instead of $\lambda=1250 \ mmd^{-1}$. It is well known that the erosion term has a significant effect in long-term simulations as it regulates the maximum biofilm thickness representing a negative rate in the free boundary Equation (\ref{d.4}) \cite{Abbas_Eberl_2012}. In the present case, the limited forward filtration time does not allow the erosion term to significantly affect biofilm growth and evolution. To generalize the problem, the term $\lambda L^{2}$ was not removed from Equation (\ref{d.5}). This can be useful to test the model on different engineering systems where longer forward filtration phases are adopted. In the present work, the results showed a similar trend of the relative errors in all the tested cases and confirmed the negligible influence of the erosion term on the detachment rate (data not shown).

Finally, other numerical simulations were performed by decreasing the initial volume fraction of the EPS component within the biofilm (from $0.05$ to $0.01 \%$). The obtained results (data not shown) showed a very similar trend of the pressure drop profile, but a better fit with experimental data was obtained when using an EPS volume fraction of $0.05$ (data not shown). This evidence could be ascribed to the presence of EPS in the wastewater. These compounds immediately contribute to increase the pressure drop during the initial phase of microfiltration. Indeed, the result confirms the negligible effect of the EPS matrix on the pressure drop during the initial phase of the experiment, and the highly relevant effect of the same component during biofilm growth and maturation.

\bibliographystyle{elsarticle-num}
\bibliography{Luongo_ArXiv}

\end{document}